%% file: main.tex
\documentclass{IEEEtran4PSCC}
\ifCLASSINFOpdf
   \usepackage[pdftex]{graphicx}
\else
   \usepackage[dvips]{graphicx}
\fi
\usepackage{makecell}

\usepackage[cmex10]{amsmath}
\usepackage{balance} 
\usepackage{subfig} 
\usepackage{multicol}
\usepackage{amssymb}
\usepackage{cite}

\usepackage{amsthm}  

\newtheorem*{remark}{Remark}

\allowdisplaybreaks

\makeatletter
\let\old@ps@headings\ps@headings
\let\old@ps@IEEEtitlepagestyle\ps@IEEEtitlepagestyle
\def\psccfooter#1{%
    \def\ps@headings{%
        \old@ps@headings%
        \def\@oddfoot{\strut\hfill#1\hfill\strut}%
        \def\@evenfoot{\strut\hfill#1\hfill\strut}%
    }%
    \def\ps@IEEEtitlepagestyle{%
        \old@ps@IEEEtitlepagestyle%
        \def\@oddfoot{\strut\hfill#1\hfill\strut}%
        \def\@evenfoot{\strut\hfill#1\hfill\strut}%
    }%
    \ps@headings%
}
\makeatother

\psccfooter{%
        \parbox{\textwidth}{\hrulefill \\ \small{24th Power Systems Computation Conference} \hfill \begin{minipage}{0.2\textwidth}\centering \vspace*{4pt} \includegraphics[scale=0.06]{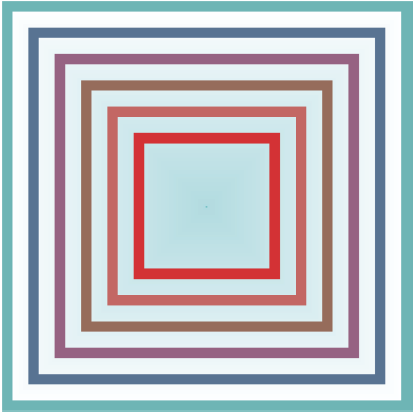}\\\small{PSCC 2026} \end{minipage} \hfill \small{Limassol, Cyprus --- June 8 -- June 12, 2026}}%
}

\begin{document}
\bstctlcite{IEEEexample:BSTcontrol}
%

\twocolumn

\title{Real-time Measurement-based Optimization for Distribution System Operation Considering\\ Battery Voltage and Thermal Constraints}


\author{\IEEEauthorblockN{Sen Zhan\IEEEauthorrefmark{1}, Lingkang Jin\IEEEauthorrefmark{1}, Haoyang Zhang\IEEEauthorrefmark{1}, Nikolaos G. Paterakis\IEEEauthorrefmark{1}}
\IEEEauthorblockA{\IEEEauthorrefmark{1}Electrical Energy Systems, Eindhoven University of Technology,
Eindhoven, The Netherlands\\ Email: \{s.zhan, l.jin, h.zhang2, n.paterakis\}@tue.nl}
}

\maketitle

\begin{abstract}

The secure operation of power distribution systems is challenged by the growing integration of distributed energy resources. Leveraging the flexibility of battery storage offers a cost-effective alternative to measures like generation curtailment, which results in energy losses. However, developing an effective operational model for battery storage is hindered by inaccurate grid models, unavailability of load data, nonlinear relationship between power injections and network states, intertemporal constraints, and complex electrochemical and thermal dynamics. To address these challenges, this paper proposes a data-driven operational control scheme for battery storage in distribution systems. Linear and convex quadratic operational constraints are constructed based on real-time distribution system and battery storage measurements. Lyapunov optimization decouples multi-period battery operation, enabling a real-time, forecast-free control strategy with low computational complexity. Numerical studies using nonlinear distribution system and battery storage simulators validate the effectiveness of the approach in ensuring secure distribution system operation and satisfaction of voltage and thermal constraints of battery storage.

\end{abstract}



 \thanksto{\noindent This work is part of the NO-GIZMOS project (MOOI52109) which received funding from the Topsector Energie MOOI subsidy program of the Netherlands Ministry of Economic Affairs and Climate Policy, executed by the Netherlands Enterprise Agency (RVO).}

\input{part1.tex}
\input{part2.tex}
\input{part3.tex}

\balance

\bibliographystyle{IEEEtran.bst}
\bibliography{ref}

\end{document}

%% file: part1.tex
\section{Introduction}

The rapid integration of distributed energy resources (DERs) introduces increasing operational challenges to distribution system operators (DSOs) in maintaining voltage and loading limits. As grid reinforcement remains a costly and time-consuming process, and requires a skilled workforce, leveraging the flexibility of DERs has emerged as a promising and cost-effective alternative. Battery storage offers both upward and downward flexibility without incurring energy losses like generation curtailment. Furthermore, as the net metering policy is being phased out \cite{RVO2025netting}, household battery storage adoption is expected to increase.

Developing an effective operational model for grid-support battery storage is challenged by several factors including: (i) inaccurate grid models, (ii) unavailability of load data, (iii) nonlinear power flow equations, (iv) intertemporal constraints, and (v) complex electrochemical and thermal dynamics. As noted in~\cite{Bassi2022}, accurate electrical models are not always readily available for distribution networks. This, combined with the lack of real-time load data and the complex power flow equations, makes it difficult to assess the impact of battery storage operation on grid voltages and loadings. Furthermore, the intertemporal nature of battery storage--where decisions at one time step affect future operation--necessitates a multi-period formulation and the use of short-term profile forecasts~\cite{Zhan2024_pscc,Zhou2020}. Finally, many existing battery storage models overlook the electrochemical and thermal dynamics \cite{Zhan2024_pscc,Zhou2020,Fan2021,Stai2021}, which can lead to violations of cell voltage and temperature constraints or produce control setpoints that are infeasible due to internal protection mechanisms. Capturing these dynamics in a tractable manner is therefore essential for reliable battery storage operation.

Model-based optimization approaches typically require detailed knowledge of distribution networks and accurate load data, which are often unavailable in practice. Moreover, they depend on the explicit modeling of nonlinear power flow equations, resulting in high computational complexity and limited scalability. To address these challenges, measurement-based optimization—also referred to as feedback optimization—has been proposed in \cite{Hauswirth2021,Gupta2025,Bernstein2019,Bolognani2019,Nowak2020,Gupta2022}. This approach replaces conventional power flow equations with measurement-based linear power flow models, where network sensitivities can be learnt from operational data \cite{Gupta2022,Gupta2025,Nowak2020}. By leveraging real-time measurements, it eliminates the need for load data and offers enhanced robustness to model uncertainties \cite{Hauswirth2021,Ortmann2020}.

The intertemporal nature of battery storage operation complicates real-time decision-making, as current decisions affect future operation. To address this, short-term forecasts and multi-period optimization formulations are commonly employed, as in \cite{Zhou2020,Zhan2024_pscc}. To reduce computational complexity and eliminate the need for forecasts, Lyapunov optimization has been explored in works such as \cite{Stai2021,Fan2021,Zhan2025_pt}. This approach augments the instantaneous cost function with a drift term, resulting in a drift-plus-penalty framework that balances performance optimality with battery state-of-charge (SoC) stability.

Many existing works neglect the electrochemical and thermal dynamics of battery storage \cite{Zhan2024_pscc,Zhou2020,Fan2021,Stai2021}, which can lead to violations of cell voltage and temperature limits. Several efforts have been made to consider this aspect. For example, in~\cite{Morstyn2020}, a second-order cone programming model was developed to capture battery voltage dynamics. In \cite{Chen2024,Lei2025}, integrated thermal and voltage models were proposed that are compatible with optimal dispatch frameworks. Nevertheless, the effectiveness of these model-based approaches hinges on precise system parameters and may suffer from error accumulation during real-time operation.

This paper aims to overcome these challenges by developing a real-time operational control scheme that explicitly considers voltage and temperature constraints of battery storage systems. The main contributions are as follows:
\begin{itemize}
  \item We propose linear and convex quadratic models to efficiently characterize the electrochemical and thermal dynamics of battery storage systems using historical operational data and real-time measurements.
  \item Leveraging Lyapunov optimization, the proposed real-time battery storage control strategy operates without accurate grid models and forecasts and generates control setpoints that balance optimality and SoC stability.
\end{itemize}

The remainder of the paper is organized as follows: \mbox{Section \ref{sec:control}} formulates the real-time control problem, introducing the Lyapunov optimization framework, network model, and battery storage model. \mbox{Section \ref{sec:case}} presents a numerical case study. \mbox{Section \ref{sec:concl}} concludes the paper and outlines directions for future work.

%% file: part2.tex
\section{Methodology} \label{sec:control}


\subsection{Notation}

Consider a distribution grid with $N+1$ buses collected in the set $\mathcal{N} = \{0,1,\cdots,N\}$, and branches including cables and transformers collected in $\mathcal{E} = \{(i,j)\} \subseteq \mathcal{N} \times \mathcal{N}$. Define $\mathcal{N}^{+} = \mathcal{N}\setminus\{0\}$, where bus 0 is the substation bus. Let $\mathcal{T}$ denote the set of time steps, and let $\Delta t$ denote the step duration. For each bus $i \in \mathcal{N}$, let $v_{i,t}$ denote its voltage magnitude at time step $t$, bounded by the lower and upper limits $\underline{v}$ and $\overline{v}$, respectively. The voltage magnitude at the substation bus is fixed at $v_0$. For each branch $(i,j) \in \mathcal{E}$, let $P_{ij,t}$ and $Q_{ij,t}$ be the active and reactive power flows from bus $i$ to $j$ at time $t$, and let $\overline{S}_{ij}$ denote the apparent power capacity of the branch.

Define $\mathcal{N}^{pv} \subseteq \mathcal{N}^{+}$ as the set of buses with PV installations, and $\mathcal{N}^{b} \subseteq \mathcal{N}^{+}$ as the set of buses with battery storage syst
ems. For each PV unit $i \in \mathcal{N}^{pv}$ and time $t$, let $p^{pv}_{i,t}$ and $q^{pv}_{i,t}$ be its active and reactive power injections into the grid (negative values for offtake), $\overline{P}^{pv}_{i,t}$ its maximum power generation, and $\overline{S}^{pv}_i$ its inverter capacity. For each battery storage unit $i \in \mathcal{N}^{b}$, let $p^b_{i,t}$ denote its charging power at time $t$ (negative values for discharging), and $q^b_{i,t}$ its reactive power consumption (negative values for injection). Let $\overline{S}^b_i$ be its power rating, $\overline{E}^b_i$ its energy capacity, and $soc_{i,t}$ its state of charge at time $t$, constrained by $\underline{soc}_i \leq soc_{i,t} \leq \overline{soc}_i$. Let $\theta_{i,t}$ denote its temperature, bounded by $\theta_{i,t} \leq \overline{\theta}_i$, and $v^{cell}_{i,t}$ its cell voltage, bounded by $\underline{v}^{cell}_i \leq v^{cell}_{i,t} \leq \overline{v}^{cell}_i$. Let $\eta^{ch}_i$ and $\eta^{dis}_i$ denote the charging and discharging efficiencies, respectively. Finally, boldface letters represent column vectors composed of the corresponding quantities, e.g. $\mathbf{v} = \left[ v_i, i \in \mathcal{N}^{+} \right]^{\top}$.

\subsection{Lyapunov optimization framework}

Multi-period optimization is a common approach for modeling and dispatching battery storage, but it requires a complex multi-period framework, limits scalability, and depends on short-term forecasts. In contrast, the objective of this paper is to design a real-time decision-making tool that operates without forecasts. To address the temporal coupling of battery storage operation, we employ a Lyapunov optimization framework, which minimizes the sum of the instantaneous cost and the Lyapunov drift, achieving a balance between operational performance, computational efficiency, and data requirements.

For each battery $i \in \mathcal{N}^b$ and time step $t \in \mathcal{T}$, define a virtual queue
\begin{equation}
    Q_{i,t} = soc_{i,t} \overline{E}_i^b - \hat{e}_i,
\end{equation}
which reflects the deviation from a reference state of energy $\hat{e}_i$. The virtual queue satisfies
\begin{equation}
    Q_{i,t+1} = Q_{i,t} + \xi^+_{i,t}- \xi^-_{i,t},
\end{equation}
where $\xi^+_{i,t}=[q_{i,t}^b]^+ \eta_{ch} \Delta t$, $\xi^-_{i,t} = [-q_{i,t}^b]^+/\eta_{dis} \Delta t$ are the incoming, outgoing energy with $[\cdot]^+=\max(\cdot,0)$. Define the Lyapunov function as
\begin{equation}
    L_t = \frac{1}{2} \sum_{i \in \mathcal{N}^b} Q_{i,t}^2,
\end{equation}
which is a scalar reflecting the total system deviation from the reference state of energy.

The one-step Lyapunov drift is
\begin{align}
   L_{t+1} - L_t &= \frac{1}{2} \sum_{i \in \mathcal{N}^b} Q_{i,t+1}^2 - \frac{1}{2} \sum_{i \in \mathcal{N}^b} Q_{i,t}^2 \nonumber\\
   &\leq \sum_{i \in \mathcal{N}^b} Q_{i,t}(\xi^+_{i,t}- \xi^-_{i,t}) + B,
\end{align}
where $B$ is a constant bounding $\frac{1}{2} \sum_{i \in \mathcal{N}^b}(\xi^+_{i,t}- \xi^-_{i,t})^2$ from above. Such a bound arises naturally from the charging/discharging rate limit.

Let $f_t$ denote the instantaneous operational cost at time $t$, which may include PV curtailment cost and (virtual) reactive power cost. The Lyapunov optimization framework seeks to minimize the \emph{drift-plus-penalty} in (\ref{eq:drift_plus_penalty}). In fact, (\ref{eq:drift_plus_penalty}) minimizes an upper bound on the Lyapunov drift, which indirectly regulates the true drift to guarantee stability while simultaneously minimizing the operational cost.
\begin{align}
    \underset{ p^b_{i,t}, q^b_{i,t},p^{pv}_{i,t}, q^{pv}_{i,t}  }{\operatorname{minimize}} &
    \quad f_t + \gamma \sum_{i\in\mathcal{N}^b} Q_{i,t}(\xi^+_{i,t}- \xi^-_{i,t}), \label{eq:drift_plus_penalty}\\
       \text{s. t. } & \underline{v} \leq v_{i,t} \leq \overline{v}, \forall i \in \mathcal{N}^+, \label{cons:voltage}\\
   & \sqrt{P^2_{ij,t}+Q^2_{ij,t}} \leq \overline{S}_{ij}, \forall (i,j) \in \mathcal{E}, \label{cons:current}\\
&  0\leq p^{pv}_{i,t} \leq \overline{P}^{pv}_{i,t}, \forall i \in \mathcal{N}^{pv},\label{eq:pvpower} \\
 (p^{pv}_{i,t})^2 &+ (q^{pv}_{i,t})^2 \leq (\overline{S}_i^{pv})^2, \forall i \in \mathcal{N}^{pv}, \label{eq:pvreactivepower} 
\end{align}
 \begin{align}
 \label{eq:powerlimit}
   &\eta_i^{dis} \overline{E}_i \frac{\underline{soc}_i -soc_{i, t}}{\Delta t} \leq  p_{i,t}^{b} \leq      \overline{E}_i \frac{\overline{soc}_i -soc_{i, t}}{ \eta_i^{ch} \Delta t}, \forall i \in \mathcal{N}^{b},\\
  &(p^{b}_{i,t})^2 + (q^{b}_{i,t})^2 \leq (\overline{S}_i^{b})^2, \forall i \in \mathcal{N}^{b}, \label{eq:bsreactivepower} \\
&   \underline{v}_{i}^{cell} \leq v_{i,t}^{cell} \leq  \overline{v}_{i}^{cell}, \forall i \in \mathcal{N}^{b},\label{cons:cellvoltage}\\
 &     \theta_{i,t} \leq  \overline{\theta}_{i}, \forall i \in \mathcal{N}^{b},\label{cons:celltemp}
\end{align}
where $\gamma > 0$ is a control parameter that trades off queue stability and operational performance. The problem is subject to distribution system voltage limits \eqref{cons:voltage}, branch loading limits \eqref{cons:current}, maximum PV generation limits \eqref{eq:pvpower}, PV inverter limits \eqref{eq:pvreactivepower}, battery SoC limits \eqref{eq:powerlimit}, battery storage inverter limits \eqref{eq:bsreactivepower}, cell voltage limits \eqref{cons:cellvoltage}, and temperature limits \eqref{cons:celltemp}. Explicit formulations of the system constraints and the cell voltage and temperature constraints will be given in the following sections.

\subsection{Distribution network model}

A distribution network model captures the impact of PV and battery storage operation on distribution system states such as voltages and branch loadings. Because conventional power flow equations are non-convex and require accurate network topology and load data, measurement-based linearized formulations are adopted in \eqref{eq:vmeas}-\eqref{eq:loadmeas}, where $\mathbf{\tilde{v}}_{t}$ and $\mathbf{\tilde{\ell}}_{t}$ are respectively the measured voltages and branch apparent power at the current time step $t$, and $\mathbf{u}_t = \big[
\{p^{pv}_{i,t}, q^{pv}_{i,t}\}_{i \in \mathcal{N}^{pv}},
\{p^{b}_{i,t}, q^{b}_{i,t}\}_{i \in \mathcal{N}^{b}}
\big]^\top$ collects active and reactive power from controllable PV and battery storage units. The time-varying matrices $\mathbf{S}_t^v$ and $ \mathbf{S}_t^\ell$ represent the voltage and branch loading sensitivities, respectively, reflecting how changes in controllable power affect the system states.
\begin{align}
    &\mathbf{v}_{t+1} =  \mathbf{\tilde{v}}_{t} + \mathbf{S}_t^v (\mathbf{u}_{t+1}-\mathbf{u}_{t}), \label{eq:vmeas}\\
    &\boldsymbol{\ell}_{t+1} =  \boldsymbol{\tilde{\ell}}_{t} + \mathbf{S}_t^\ell (\mathbf{u}_{t+1}-\mathbf{u}_{t}).\label{eq:loadmeas}
\end{align}

\begin{figure}[t]
  \centering
  \subfloat[Voltage]{\includegraphics[width=0.8\columnwidth]{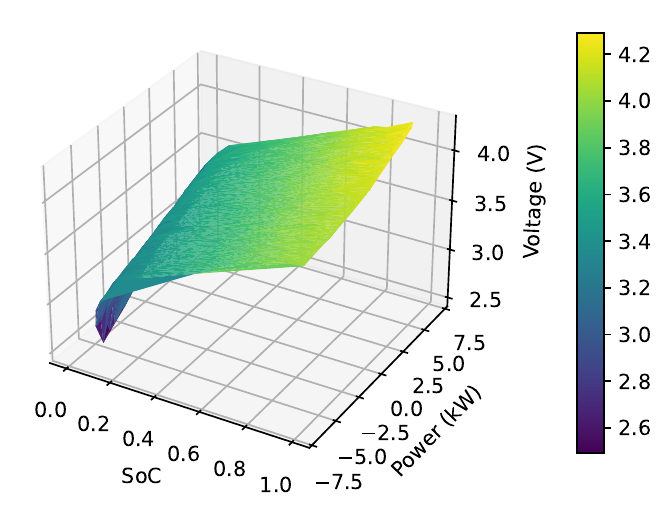}}\\
  \vspace{-10pt}
  \subfloat[Temperature]{\includegraphics[width=0.8\columnwidth]{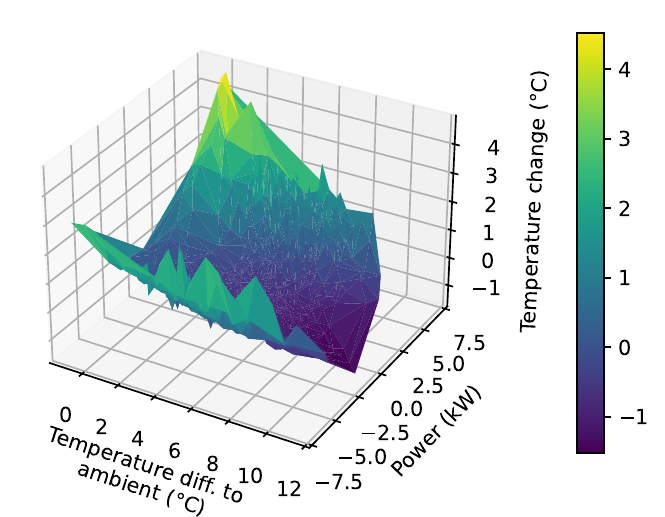}}
  \caption{Voltage and temperature profiles based on simulated operational data for a 6.8 kW (0.5C) lithium-ion NMC battery using the PyBaMM simulator \cite{pybamm}. The battery is cycled between low and high SoC limits, with charging and discharging power setpoints randomly selected at each time step. Data points are collected every 5 minutes.}
  \label{fig:3dplot}
\end{figure}

\subsection{Battery storage model}
Likewise, exact modeling of the electrochemical and thermal dynamics of battery storage is complex and requires identifying key system parameters \cite{Lei2025}. Furthermore, model-based approaches typically lack robustness \cite{Hauswirth2021}, where approximation errors can accumulate during operation and lead to cell voltage and temperature limit violations. In this regard, we aim to leverage the robustness of measurement-based approaches and extend them to battery storage modeling. 

Figure \ref{fig:3dplot} presents the simulated voltage and temperature profiles of a 6.8 kW (0.5C) lithium-ion NMC battery, obtained using the specialized differential equation solver PyBaMM~\cite{pybamm}. In Fig. \ref{fig:3dplot}a, the cell voltages exhibit an approximately linear relationship with power, with the slope strongly influenced by the state of charge. Figure \ref{fig:3dplot}b shows that the cell temperature dynamics are approximately linear in the temperature difference to the ambient and quadratic in power. Motivated by these observations, we propose the measurement-based models \eqref{eq:voltmeas}-\eqref{eq:tempmeas} to characterize cell voltages and temperature. In \eqref{eq:voltmeas}, $\boldsymbol{\Lambda}_t^{v}$ is a diagonal matrix whose elements map power to cell voltage variations. Similarly, in \eqref{eq:tempmeas}, the diagonal matrices $\boldsymbol{\Lambda}_t^{\theta}$ and $\boldsymbol{\Lambda}_t^{amb}$ capture the influence of quadratic power and ambient temperature difference on cell temperature, respectively, where~$\circ$ denotes the element-wise multiplication of vectors.
\begin{align}
   & \mathbf{v}^{cell}_{t+1} =  \mathbf{\tilde{v}}^{cell}_{t} + \boldsymbol{\Lambda}_t^{v} \mathbf{p}^b_{t+1}, \label{eq:voltmeas}\\
   & \boldsymbol{\theta}_{t+1} =  \boldsymbol{\tilde{\theta}}_{t} + \boldsymbol{\Lambda}_t^{\theta} \mathbf{p}^b_{t+1} \circ \mathbf{p}^b_{t+1} + \boldsymbol{\Lambda}_t^{amb} (\boldsymbol{\tilde{\theta}}_{t}-\boldsymbol{\tilde{\theta}}^{amb}_{t}). \label{eq:tempmeas}
\end{align}

\subsection{Solution process}
Based on the distribution network and battery storage models, we aim to develop an iterative algorithm that tracks the time-varying optimizer of \eqref{eq:drift_plus_penalty}-\eqref{cons:celltemp} at each time step using only current system measurements without any forecasts. Particularly, we adopt a projected gradient descent (PGD) algorithm from \cite{Haberle2021,Hauswirth2021} that involves solving the following optimization problem at each time step $t$.
\begin{align} 
\underset{\mathbf{u}}{\operatorname{minimize}}  & \quad \| \mathbf{u} - (\mathbf{u}_{t} - \alpha \nabla_{\mathbf{u}}  g_t|\mathbf{u}_{t}) \|^2, \label{eq:objgrad} \\
\text{s. t. } & \mathbf{u}  = \big[
\{p^{pv}_{i}, q^{pv}_{i}\}_{i \in \mathcal{N}^{pv}},
\{p^{b}_{i}, q^{b}_{i}\}_{i \in \mathcal{N}^{b}}
\big]^\top, \label{eq:ugrad}  \\
& \mathbf{\underline{v}} \leq  \mathbf{\tilde{v}}_{t} + \mathbf{S}_t^v (\mathbf{u}-\mathbf{u}_{t}) \leq \mathbf{\overline{v}}, \label{eq:voltgrad} \\
& \boldsymbol{\tilde{\ell}}_{t} + \mathbf{S}_t^\ell (\mathbf{u}-\mathbf{u}_{t}) \leq [\overline{S}_{ij}, \forall (i,j)\in\mathcal{E}]^\top, \label{eq:linegrad} \\
&  0\leq p^{pv}_{i} \leq \overline{P}^{pv}_{i,t}, \forall i \in \mathcal{N}^{pv},\label{eq:pvpowergrad} \\
 (p^{pv}_{i})^2 &+ (q^{pv}_{i})^2 \leq (\overline{S}_i^{pv})^2, \forall i \in \mathcal{N}^{pv}, \label{eq:pvreactivepowergrad} \\
 \label{eq:powerlimitgrad}
   \eta_i^{dis} \overline{E}_i &\frac{\underline{soc}_i -soc_{i, t}}{\Delta t} \leq p_{i}^{b} \leq      \overline{E}_i \frac{\overline{soc}_i -soc_{i, t}}{ \eta_i^{ch} \Delta t}, \\
  (p^{b}_{i})^2 &+ (q^{b}_{i})^2 \leq (\overline{S}_i^{b})^2, \forall i \in \mathcal{N}^{b}, \label{eq:bsreactivepowergrad}\\
     & \mathbf{\underline{v}}^{cell} \leq  \mathbf{\tilde{v}}^{cell}_{t} + \boldsymbol{\Lambda}_t^{v} \mathbf{p}^b \leq \mathbf{\overline{v}}^{cell}, \label{eq:voltmeasgrad}\\
   &  \boldsymbol{\tilde{\theta}}_{t} + \boldsymbol{\Lambda}_t^{\theta} \mathbf{p}^b \circ \mathbf{p}^b + \boldsymbol{\Lambda}_t^{amb} (\boldsymbol{\tilde{\theta}}_{t}-\boldsymbol{\tilde{\theta}}^{amb}_{t}) \leq \boldsymbol{\overline{\theta}}. \label{eq:tempmeasgrad}
\end{align}

The objective function in \eqref{eq:objgrad} represents the Euclidean projection of the gradient descent iterate, where $g_t = f_t + \gamma \sum_{i\in\mathcal{N}^b} Q_{i,t}(\xi^+_{i,t}- \xi^-_{i,t})$ is the objective function in \eqref{eq:drift_plus_penalty} and $\alpha$ is the step size. In other words, the update seeks the closest feasible point to the unconstrained gradient step, thereby ensuring that the control decisions respect the operational limits of the system. The decision vector $\mathbf{u}$ in \eqref{eq:ugrad}
collects active and reactive power setpoints of PV inverters and batteries. Constraints \eqref{eq:voltgrad}-\eqref{eq:linegrad} enforce linearized voltage and line limits. Individual DER limits are defined in \eqref{eq:pvpowergrad}-\eqref{eq:bsreactivepowergrad}. Local 
measurement-based cell voltage limits are incorporated in \eqref{eq:voltmeasgrad}, and thermal limits accounting for quadratic power dependence and ambient 
temperature are imposed in \eqref{eq:tempmeasgrad}. 

At each time step $t$, after collecting the required parameters, the optimization problem is solved. The resulting solution $\mathbf{u}^*$ is then dispatched, i.e. the next iterate is updated as $\mathbf{u}_{t+1} \gets \mathbf{u}^*$. Note that this approach requires a centralized communication architecture. Distributed implementations based on primal-dual methods have been proposed in~\cite{Zhan2025_pt}, while the PGD algorithm offers improved convergence properties \cite{Haberle2021}.

\subsection{Convergence analysis}\label{sec:proof}

We establish the convergence of the proposed algorithm under the following assumptions:  
\begin{enumerate}
    \item The batteries are assumed to be lossless.  
    \item The cost function $f_t$ is convex, continuously differentiable, and has Lipschitz continuous gradient with Lipschitz constant $L > 0$.  
    \item The approximations \eqref{eq:vmeas}--\eqref{eq:loadmeas} and \eqref{eq:voltmeas}--\eqref{eq:tempmeas} are exact.  
\end{enumerate}

Under Assumption 1, the second term in the cost function is linear in $\mathbf{p}^b_t$. Together with Assumption 2, it follows that the objective function in \eqref{eq:drift_plus_penalty} is convex and possesses a Lipschitz continuous gradient with constant $L$. By Assumption~3, the optimization problem defined by \eqref{eq:drift_plus_penalty}--\eqref{cons:celltemp} is a convex quadratically constrained quadratic program (QCQP). According to Proposition 2.3.2 in \cite{bertsekas1999nonlinear}, the projected gradient descent method specified in \eqref{eq:objgrad}--\eqref{eq:tempmeasgrad} converges to a minimizer of \eqref{eq:drift_plus_penalty}--\eqref{cons:celltemp} if $0 < \alpha < 2/L$.  

\begin{remark}
The above convergence result relies on the assumptions of lossless batteries and exact grid and battery model approximations. In the following simulation study, we evaluate the algorithm using nonlinear grid and battery simulators that account for battery losses and model nonlinearities.  
\end{remark}

\subsection{Sensitivity matrices}
The sensitivity matrices $\mathbf{S}_t^v$, $\mathbf{S}_t^\ell$, $\boldsymbol{\Lambda}_t^v$, $\boldsymbol{\Lambda}_t^\theta$, and $\boldsymbol{\Lambda}_t^{amb}$ play a central role in the algorithm. While the feedback-based design and the adjustable step size $\alpha$ allow constraint satisfaction even with inaccurate sensitivities \cite{Zhan2025_robust}, accurate values are essential for achieving near-optimal performance. To estimate the distribution system sensitivities $\mathbf{S}_t^v$ and $\mathbf{S}_t^\ell$, the linearized \textit{DistFlow} model \cite{Farivar2013} can be employed using topology data and line parameters. Alternative approaches include the perturb-and-observe method for general network topologies reported in \cite{Fu2022}, and a regression-based method leveraging historical operational data presented in \cite{Picallo2022}. Since batteries exhibit an approximately linear relationship between cell voltage and power—where the slope depends on the SoC—we propose the following robust approximation method \eqref{eq:slopeapprox}, where $\lambda_{i,t}^v$ denotes the $i$-th diagonal element of $\mathbf{S}t^v$. Specifically, the slope is selected as the maximum observed slope in historical data corresponding to SoC values within a neighborhood of $soc_{i,t}$, where the neighborhood is defined by $\sigma>0$: 
\begin{align}
    & \lambda_{i,t}^v = \max_{k|soc_{i,k} - \sigma \leq soc_{i,t} \leq soc_{i,k} + \sigma } \frac{v^{cell}_{i,k+1}-v^{cell}_{i,k}}{p^b_{i,k+1}}. \label{eq:slopeapprox}
\end{align}
Finally, the thermal model sensitivities were obtained via linear regression on historical data.

%% file: part3.tex
\section{Numerical Validation}\label{sec:case}
 

In this section, we present a numerical study to evaluate the effectiveness of the proposed strategy in addressing voltage and congestion management as well as battery operational constraints.  Although the optimization is based on linear, measurement-driven models, its validation is conducted through detailed nonlinear simulations to ensure real-world applicability. Battery behavior is simulated using \textit{PyBaMM}~\cite{pybamm}, which captures nonlinear electrochemical and thermal dynamics, whereas the distribution system is modeled with the AC power flow solver \textit{power-grid-model} \cite{Xiang2023,Xiang_PowerGridModel_power-grid-model}.

\subsection{Case description}
\begin{figure}[t]
    \centering
    \includegraphics[width=\columnwidth]{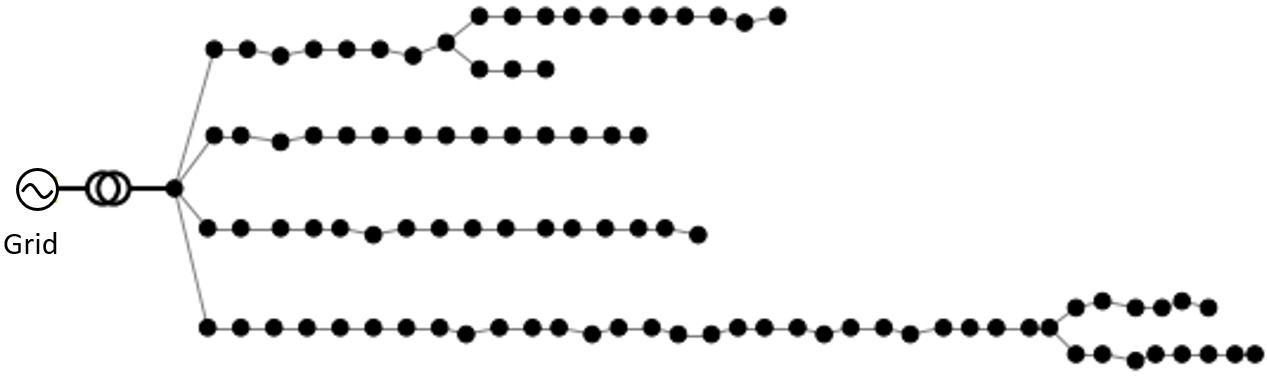} 
    \caption{A 97-bus test grid adopted from \textit{Simbench} \cite{Meinecke2020SimBench}.}
    \label{fig:grid}
\end{figure}

The simulations are carried out on a 97-bus low-voltage grid from \textit{SimBench} \cite{Meinecke2020SimBench}, illustrated in Fig.~\ref{fig:grid}. To emulate a congested scenario for demonstration, 54 PV units and 36 battery storage systems are randomly placed across the network, while the transformer capacity is increased from 250 to 400 kVA. Time-series simulations are performed using data from a representative summer day in the original \textit{SimBench} dataset. The optimization algorithm is executed every five minutes, providing updated setpoints for the PV and battery storage units. The operational limits are set as follows: the distribution system voltages are between 0.95 and 1.05 pu; the battery charging and discharging efficiencies are 0.97; the initial SoC is set to 0.0; the minimum and maximum cell voltages are 2.5 and 4.2 V; and the upper cell temperature limit is 45$^{\circ}$C. The cost function is defined as  $f_t = \frac{1}{2} \sum_{i \in \mathcal{N}^{pv}} ( p_{i,t}^{pv} - \overline{P}_{i,t}^{pv} )^{2} + \frac{1}{2}\omega \big[ \sum_{i \in \mathcal{N}^{pv}} (q_{i,t}^{pv})^2 +  \sum_{i \in \mathcal{N}^{b}} (p_{i,t}^{ch})^2
  +  \sum_{i \in \mathcal{N}^{b}} (q_{i,t}^{ch})^2 \big]$
with $\omega=0.1$, implying that PV curtailment is penalized more heavily than other sources of flexibility. The parameter $\hat{e}_i$ is set to 0.0 to maximize the charging capability for excess PV generation, and $\sigma$ in \eqref{eq:slopeapprox} is set to 0.1. The Lyapunov drift parameter $\gamma$ is set to 0.05, and the step size $\alpha$ to 0.5, resulting in stable numerical performance. The impact of various $\gamma$ and $\alpha$ values is examined in Section \ref{sec:sensitivities}.

\subsection{Result}
The results are presented in four parts. First, we present regression results for cell temperature dynamics. Second, we show that neglecting battery dynamics can lead to violations of operational constraints, thereby demonstrating the necessity of incorporating them. Third, we evaluate the capability of the proposed approach to maintain voltage and transformer loading limits through real-time optimal battery dispatch. Finally, we analyze the impact of some key parameters on system performance.

\begin{figure}[t]
    \centering
    \includegraphics[width=0.8\columnwidth]{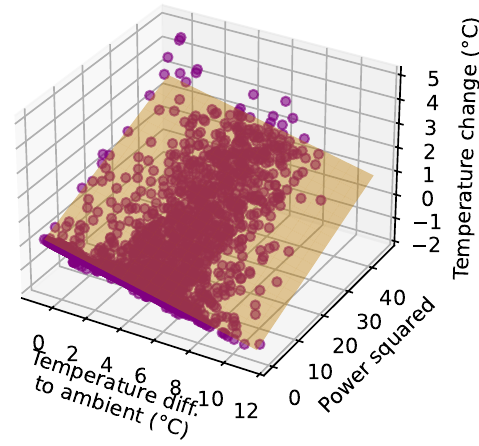} 
    \caption{A 3D representation of the linear regression model for the cell temperature dynamics of a 6.8 kW battery unit.}
    \label{fig:reg}
\end{figure}

\begin{table}[t]
\centering
\caption{Cell temperature dynamics regression results.}
\begin{tabular}{ccccc}
\hline
\makecell{Battery rating\\(kw, 0.5C)} & \makecell{Coeffients\\(~ / °C/$kW^2$)} & \makecell{MAE\\(°C)} & \makecell{RMSE\\(°C)} & \makecell{Cross-validated\\$R^2$} \\
\hline
1.7 & -0.15 / 0.92 & 0.12 & 0.27 & 0.88 ± 0.013  \\
2.9 & -0.15 / 0.32 & 0.12 & 0.28 & 0.87 ± 0.015  \\
3.4 & -0.15 / 0.23 & 0.12 & 0.25 & 0.90 ± 0.016  \\
4.3 & -0.15 / 0.14 & 0.12 & 0.26 & 0.89 ± 0.016  \\
6.8 & -0.15 / 0.06 & 0.12 & 0.25 & 0.90 ± 0.017  \\
32.5 & -0.15 / 0.003 & 0.12 & 0.26 & 0.89 ± 0.009  \\
34.6 & -0.15 / 0.002 & 0.13 & 0.27 & 0.88 ± 0.011  \\
\hline
\end{tabular}
\label{tab:temp}
\end{table}

\subsubsection{Cell temperature dynamics regression analysis} For modeling cell temperature dynamics, we propose using the temperature difference with respect to the ambient and the squared power as predictors, as expressed in \eqref{eq:tempmeas}. We validate this formulation through regression analysis. Figure~\ref{fig:reg} shows the 3D surface fitted by the linear regression model for the representative 6.8 kW battery unit. The model captures the dependence of the cell temperature rise on the cell-ambient temperature difference and the squared power.

Table~\ref{tab:temp} reports the regression results for all tested batteries. The coefficient associated with the temperature difference remains stable at approximately -0.15°C/kW², whereas the coefficient of the squared power decreases with increasing battery size, reflecting the reduced specific thermal stress in larger systems.
Across all units, the prediction accuracy is consistently high, with the mean absolute errors (MAE) below 0.13°C and the root mean square error (RMSE) of 0.25-0.28°C.
The cross-validated coefficient of determination $R^2$ remains above 0.87, indicating that the proposed model explains the observed temperature variations well.
These results confirm the effectiveness of the simplified linear model across different battery capacities and operating conditions.

\subsubsection{Incorporating battery dynamics}

\begin{figure}[t]
  \centering
  \subfloat[Voltage]{\includegraphics[width=\columnwidth]{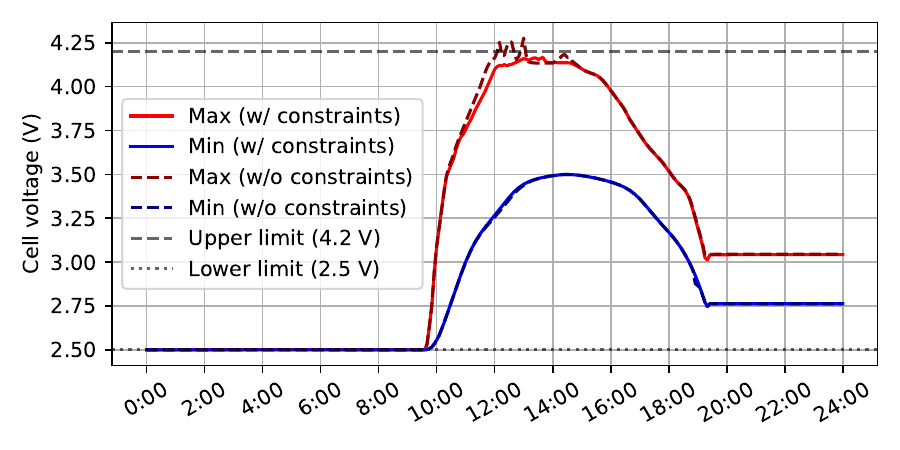}}\\
  \subfloat[Temperature]{\includegraphics[width=\columnwidth]{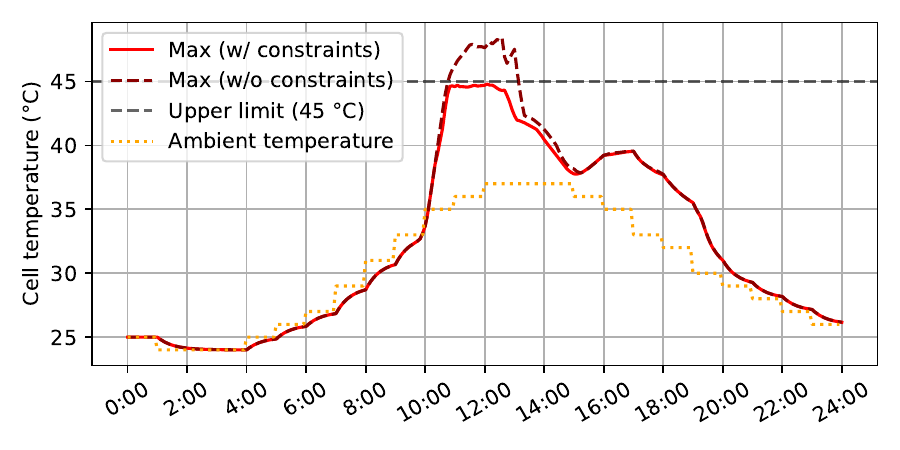}}
  \caption{Cell voltage and temperature profiles with/without measurement-based constraints.}
  \label{fig:compare}
\end{figure}

Figure \ref{fig:compare}a illustrates the maximum and minimum cell voltages with and without measurement-based constraints. The cell voltages are bounded by an upper limit of 4.2 V and a lower limit of 2.5 V. Around midday, the voltages rise significantly as the batteries charge to absorb excess PV generation. With measurement-based constraints, the voltages remain within the safety limits, while without them, the upper limit is exceeded. Similarly, Fig. \ref{fig:compare}b shows the maximum cell temperature with and without measurement-based constraints, along with the upper temperature limit of 45°C and the ambient temperature. The temperature peaks around midday. With measurement-based constraints, the temperature stays below the safety limit, whereas without them it slightly exceeds the limit during peak periods. These results highlight the importance of considering both electrochemical and thermal dynamics in a battery model: neglecting them can lead to elevated temperatures and voltage limit violations, while the measurement-based strategy keeps both voltages and temperature within safe operating limits.

\subsubsection{Network constraint satisfaction}

\begin{figure}[t]
  \centering
  \subfloat[Network voltage]{\includegraphics[width=\columnwidth]{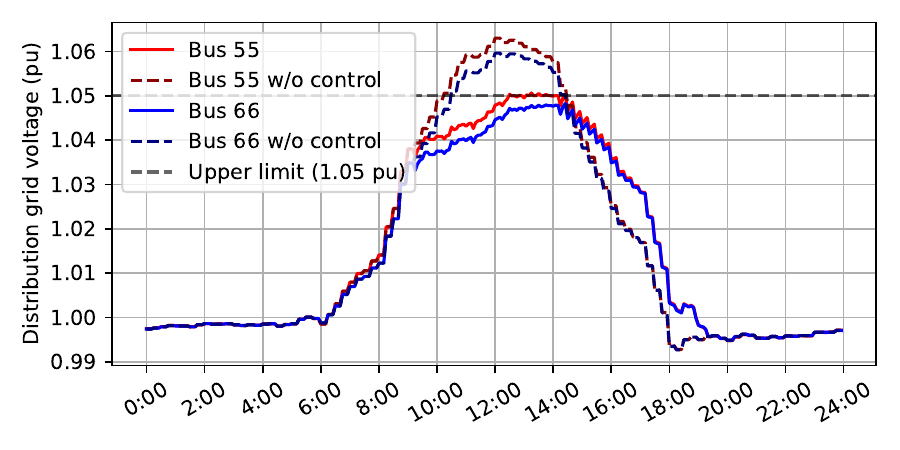}}\\
  \subfloat[Loading]{\includegraphics[width=\columnwidth]{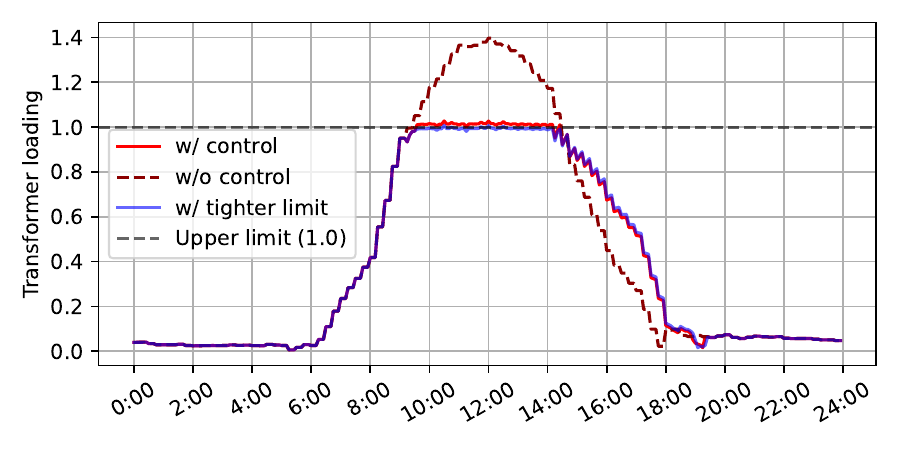}}\\
   \subfloat[Power profiles]{\includegraphics[width=\columnwidth]{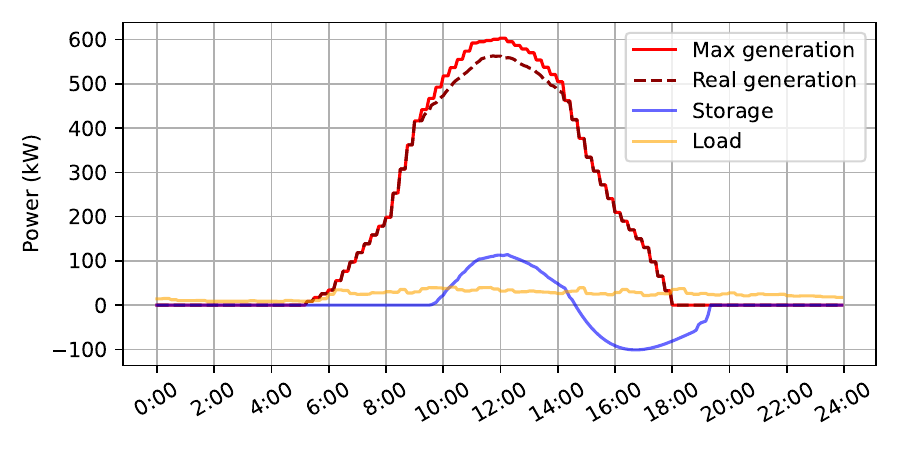}} 
  \caption{Network voltage and transformer loading with and without control and aggregated power profiles.}
  \label{fig:network}
\end{figure}

To demonstrate the effectiveness of the proposed strategy for network voltage and loading limit management, Figs. \ref{fig:network}a and \ref{fig:network}b present the bus voltage and transformer loading profiles with and without the control strategy, respectively. The results show that the proposed approach successfully maintains both the voltage and transformer loading within their limits. Minor and temporary violations of voltage and loading constraints are observed, which can be attributed to the influence of non-controllable elements in the network, such as uncontrollable load units. To further reduce these transient violations, tighter voltage and loading limits can be imposed in the model. This is illustrated in Fig. \ref{fig:network}b, where as an example setting the transformer loading limit to 0.98 results in reduced temporary constraint violations. Figure \ref{fig:network}c depicts the aggregated power profiles of PV generation, storage, and loads. During periods of high PV output, generation is curtailed and the storage units absorb excess power to satisfy grid operating constraints. After the generation peak, the battery discharges following the Lyapunov drift-based control policy, which preserves sufficient capacity to accommodate subsequent charging events in the following day.

\begin{figure}[t]
  \centering
  \subfloat[Storage]{\includegraphics[width=\columnwidth]{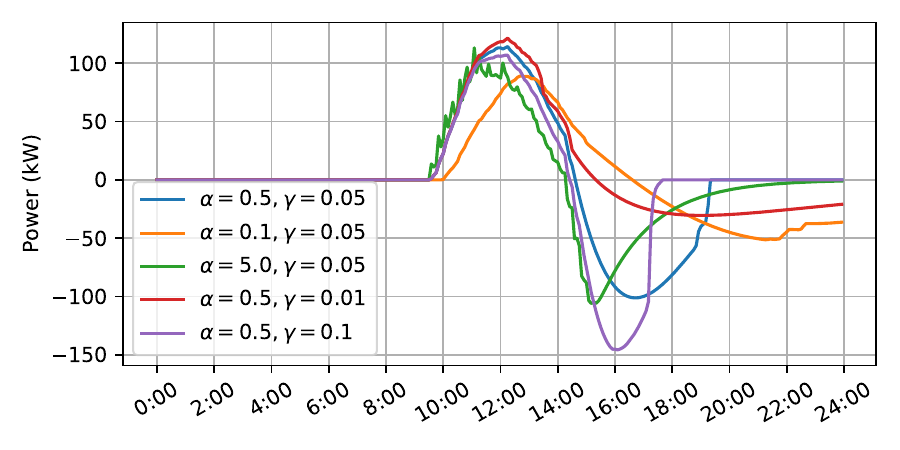}}\\
   \subfloat[PV]{\includegraphics[width=\columnwidth]{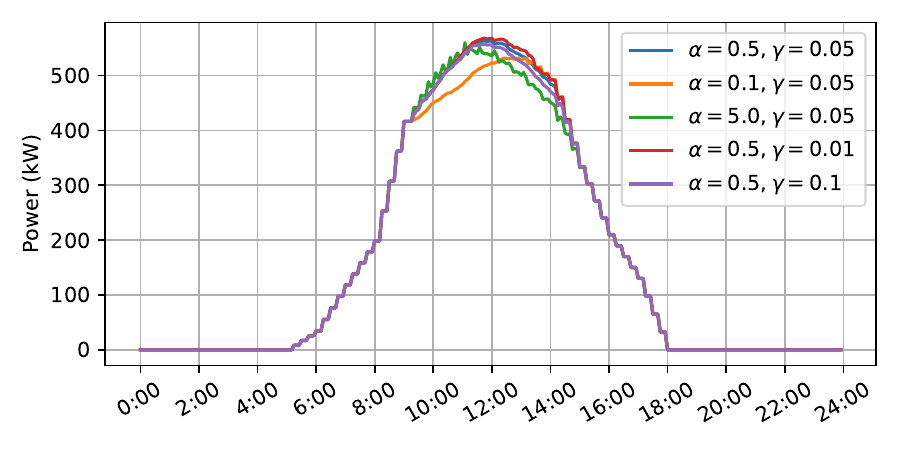}} 
  \caption{Aggregated battery storage and PV profiles under different values of $\alpha$ and $\gamma$.}
  \label{fig:sensitivity}
\end{figure}

\subsubsection{Analyzing impact of the step size and drift parameter}\label{sec:sensitivities}
This section analyzes the influence of two key parameters: the step size $\alpha$ and the drift parameter $\gamma$. Figures \ref{fig:sensitivity}a and \ref{fig:sensitivity}b illustrate the battery and PV profiles under different values of $\alpha$ and $\gamma$. A small step size slows down convergence and results in suboptimal real-time performance, as evidenced by the significantly reduced PV generation. Conversely, an excessively large step size (e.g. $\alpha = 5.0$) can cause instability. Similarly, as shown in Fig.~\ref{fig:sensitivity}a, a small drift parameter leads to a prolonged energy release process, which can limit the capacity for power absorption for the following events. In contrast, a large drift parameter allows the drift term to dominate the cost function, compromising optimality, shown by reduced PV generation. Selecting appropriate parameter values is therefore critical.  Theoretical analysis such as the step size rule in Section~\ref{sec:proof} can be used as guidelines. Fortunately, because the algorithm involves only a few parameters, tuning is relatively straightforward. Moreover, the algorithm demonstrates robust performance across a broad range of parameter settings, likely due to its feedback-based design.

\section{Conclusion} \label{sec:concl}

In this work, we have proposed a real-time, measurement-based optimization framework for distribution system operation that explicitly incorporates battery voltage and thermal constraints. By leveraging Lyapunov optimization, the method eliminates the reliance on forecasts and complex multi-period formulations, while maintaining computational tractability and robustness to model inaccuracies. The validation on a nonlinear distribution grid and detailed battery simulator demonstrates the effectiveness of the proposed approach in ensuring network security through coordinated real-time control and respecting thermal and voltage limits of batteries.

Several directions for future work remain. First, the simulations were performed under the assumption of identical battery cells. The validation focused on individual cell-level dynamics rather than full battery packs. Extending the simulation to heterogeneous battery packs is an important direction for future work. Second, while the Lyapunov-based approach provides a forecast-free solution, it inherently operates myopically and does not exploit information about expected future system states. Future research could explore incorporating the measurement-based battery storage models into the model predictive control (MPC) framework. Finally, the proposed control strategy assumes a centralized architecture with full measurement and communication availability. In real-world deployments, communication delays and data losses could degrade performance. Assessing their impact and investigating robust and distributed implementations represent another relevant extension.

%% file: main.bbl
\begin{thebibliography}{10}
\providecommand{\url}[1]{#1}
\csname url@samestyle\endcsname
\providecommand{\newblock}{\relax}
\providecommand{\bibinfo}[2]{#2}
\providecommand{\BIBentrySTDinterwordspacing}{\spaceskip=0pt\relax}
\providecommand{\BIBentryALTinterwordstretchfactor}{4}
\providecommand{\BIBentryALTinterwordspacing}{\spaceskip=\fontdimen2\font plus
\BIBentryALTinterwordstretchfactor\fontdimen3\font minus
  \fontdimen4\font\relax}
\providecommand{\BIBforeignlanguage}[2]{{%
\expandafter\ifx\csname l@#1\endcsname\relax
\typeout{** WARNING: IEEEtran.bst: No hyphenation pattern has been}%
\typeout{** loaded for the language `#1'. Using the pattern for}%
\typeout{** the default language instead.}%
\else
\language=\csname l@#1\endcsname
\fi
#2}}
\providecommand{\BIBdecl}{\relax}
\BIBdecl

\bibitem{RVO2025netting}
{Netherlands Enterprise Agency (RVO)}, ``Netting scheme for solar panels ends
  per 2027,''
  \url{https://business.gov.nl/amendment/netting-scheme-solar-panels-ends/},
  2025.

\bibitem{Bassi2022}
V.~Bassi, L.~F. Ochoa, T.~Alpcan, and C.~Leckie, ``Electrical model-free
  voltage calculations using neural networks and smart meter data,'' \emph{IEEE
  Transactions on Smart Grid}, vol.~14, pp. 3271--3282, 7 2023.

\bibitem{Zhan2024_pscc}
S.~Zhan, J.~Morren, W.~van~den Akker, A.~van~der Molen, N.~G. Paterakis, and
  J.~G. Slootweg, ``Multi-timescale coordinated distributed energy resource
  control combining local and online feedback optimization,'' \emph{Electric
  Power Systems Research}, vol. 234, 2024.

\bibitem{Zhou2020}
X.~Zhou, E.~Dall'anese, and L.~Chen, ``Online stochastic optimization of
  networked distributed energy resources,'' \emph{IEEE Transactions on
  Automatic Control}, vol.~65, pp. 2387--2401, 2020.

\bibitem{Fan2021}
S.~Fan, G.~He, X.~Zhou, and M.~Cui, ``Online optimization for networked
  distributed energy resources with time-coupling constraints,'' \emph{IEEE
  Transactions on Smart Grid}, vol.~12, pp. 251--267, 2021.

\bibitem{Stai2021}
E.~Stai, C.~Wang, and J.~Y.~L. Boudec, ``Online battery storage management via
  lyapunov optimization in active distribution grids,'' \emph{IEEE Transactions
  on Control Systems Technology}, vol.~29, pp. 672--690, 2021.

\bibitem{Hauswirth2021}
A.~Hauswirth, Z.~He, S.~Bolognani, G.~Hug, and F.~Dörfler, ``Optimization
  algorithms as robust feedback controllers,'' \emph{Annual Reviews in
  Control}, vol.~57, 2024.

\bibitem{Gupta2025}
R.~K. Gupta, P.~A. Pegoraro, O.~Stanojev, A.~Abur, C.~Muscas, G.~Hug
  \emph{et~al.}, ``Learning power flow models and constraints from
  time-synchronized measurements: A review,'' 2025.

\bibitem{Bernstein2019}
A.~Bernstein and E.~Dall'Anese, ``Real-time feedback-based optimization of
  distribution grids: A unified approach,'' \emph{IEEE Transactions on Control
  of Network Systems}, vol.~6, pp. 1197--1209, 2019.

\bibitem{Bolognani2019}
S.~Bolognani, R.~Carli, G.~Cavraro, and S.~Zampieri, ``On the need for
  communication for voltage regulation of power distribution grids,''
  \emph{IEEE Transactions on Control of Network Systems}, vol.~6, pp.
  1111--1123, 2019.

\bibitem{Nowak2020}
S.~Nowak, Y.~C. Chen, and L.~Wang, ``Measurement-based optimal der dispatch
  with a recursively estimated sensitivity model,'' \emph{IEEE Transactions on
  Power Systems}, vol.~35, pp. 4792--4802, 2020.

\bibitem{Gupta2022}
R.~Gupta, F.~Sossan, and M.~Paolone, ``Model-less robust voltage control in
  active distribution networks using sensitivity coefficients estimated from
  measurements,'' \emph{Electric Power Systems Research}, vol. 212, p. 108547,
  2022.

\bibitem{Ortmann2020}
L.~Ortmann, A.~Hauswirth, I.~Cadu, F.~Dörfler, and S.~Bolognani,
  ``Experimental validation of feedback optimization in power distribution
  grids,'' \emph{Electric Power Systems Research}, vol. 189, 2020.

\bibitem{Zhan2025_pt}
S.~Zhan, H.~Zhang, K.~Kok, and N.~G. Paterakis, ``Model-free approaches for
  real-time distribution system operation: A comparison of feedback
  optimization and reinforcement learning,'' in \emph{IEEE Powertech Kiel},
  2025.

\bibitem{Morstyn2020}
T.~Morstyn, C.~Crozier, M.~Deakin, and M.~D. McCulloch, ``Conic optimization
  for electric vehicle station smart charging with battery voltage
  constraints,'' \emph{IEEE Transactions on Transportation Electrification},
  vol.~6, pp. 478--487, 6 2020.

\bibitem{Chen2024}
Y.~Chen, K.~Zheng, Y.~Gu, J.~Wang, and Q.~Chen, ``Optimal energy dispatch of
  grid-connected electric vehicle considering lithium battery electrochemical
  model,'' \emph{IEEE Transactions on Smart Grid}, vol.~15, pp. 3000--3015, 5
  2024.

\bibitem{Lei2025}
Y.~C.~C. Chao~Lei, Shuangqi~Li, ``Optimal ship-to-grid dispatch considering
  battery thermal and voltage electrochemical-thermal-coupled constraints,''
  \emph{IEEE Transactions on Smart Grid}, 2025.

\bibitem{pybamm}
V.~Sulzer, S.~G. Marquis, R.~Timms, M.~Robinson, and S.~J. Chapman, ``Python
  battery mathematical modelling ({PyBaMM}),'' \emph{Journal of Open Research
  Software}, vol.~9, pp. 1--8, 2021.

\bibitem{Haberle2021}
V.~Haberle, A.~Hauswirth, L.~Ortmann, S.~Bolognani, and F.~Dorfler,
  ``Non-convex feedback optimization with input and output constraints,''
  \emph{IEEE Control Systems Letters}, vol.~5, pp. 343--348, 2021.

\bibitem{bertsekas1999nonlinear}
D.~P. Bertsekas, \emph{Nonlinear Programming}, 2nd~ed.\hskip 1em plus 0.5em
  minus 0.4em\relax Belmont, MA: Athena Scientific, 1999.

\bibitem{Zhan2025_robust}
S.~Zhan, J.~Morren, W.~van~den Akker, A.~van~der Molen, N.~G. Paterakis, and
  J.~G. Slootweg, ``Robustness assessment of primal-dual gradient
  projection-based online feedback optimization for real-time distribution grid
  management,'' \emph{Electric Power Systems Research}, vol. 242, 5 2025.

\bibitem{Farivar2013}
M.~Farivar, L.~Chen, and S.~Low, ``Equilibrium and dynamics of local voltage
  control in distribution systems,'' in \emph{EEE Conference on Decision and
  Control}.\hskip 1em plus 0.5em minus 0.4em\relax IEEE, 2013, pp. 4329--4334.

\bibitem{Fu2022}
A.~Fu, M.~Cvetkovic, and P.~Palensky, ``Distributed cooperation for voltage
  regulation in future distribution networks,'' \emph{IEEE Transactions on
  Smart Grid}, 2022.

\bibitem{Picallo2022}
M.~Picallo, L.~Ortmann, S.~Bolognani, and F.~Dörfler, ``Adaptive real-time
  grid operation via online feedback optimization with sensitivity
  estimation,'' \emph{Electric Power Systems Research}, vol. 212, p. 108405,
  2022.

\bibitem{Xiang2023}
Y.~Xiang, P.~Salemink, B.~Stoeller, N.~Bharambe, and W.~van Westering, ``Power
  grid model: a high-performance distribution grid calculation library,'' in
  \emph{27th International Conference on Electricity Distribution (CIRED
  2023)}, vol. 2023, 2023, pp. 1089--1093.

\bibitem{Xiang_PowerGridModel_power-grid-model}
\BIBentryALTinterwordspacing
Y.~Xiang, P.~Salemink, W.~van Westering, N.~Bharambe, M.~G. Govers, J.~van~den
  Bogaard \emph{et~al.}, ``{PowerGridModel/power-grid-model}.'' [Online].
  Available: \url{https://github.com/PowerGridModel/power-grid-model}
\BIBentrySTDinterwordspacing

\bibitem{Meinecke2020SimBench}
S.~Meinecke, D.~Sarajlić, S.~R. Drauz, A.~Klettke, L.-P. Lauven, C.~Rehtanz
  \emph{et~al.}, ``Simbench—a benchmark dataset of electric power systems to
  compare innovative solutions based on power flow analysis,'' \emph{Energies},
  vol.~13, no.~12, p. 3290, 2020.

\end{thebibliography}
